\documentclass[times, 10pt,twocolumn]{article} 
\usepackage{latex8}
\usepackage{times}
\usepackage{epsfig}

\newcommand{\ie}{{\it i.e.,$\ $}}

\begin{document}
\title{A Byzantine Fault Tolerant Distributed Commit Protocol
\thanks{This work was supported in part by Department of Energy 
Contract DE-FC26-06NT42853, and by a Faculty Research Development award
from Cleveland State University.}
 }

\author{Wenbing Zhao\\
Department of Electrical and Computer Engineering\\ 
Cleveland State University, 2121 Euclid Ave, Cleveland, OH 44115\\
wenbing@ieee.org\\
}

\maketitle
\thispagestyle{empty}

\begin{abstract}
In this paper, we present a Byzantine fault tolerant distributed commit 
protocol for transactions running over untrusted networks. The traditional 
two-phase commit protocol is enhanced by replicating the coordinator and 
by running a Byzantine agreement algorithm among the coordinator replicas.
Our protocol can tolerate Byzantine faults at the coordinator replicas and 
a subset of malicious faults at the participants. A decision certificate, 
which includes a set of registration records and a set of votes from 
participants, is used to facilitate the coordinator replicas to reach a 
Byzantine agreement on the outcome of each transaction. The certificate also 
limits the ways a faulty replica can use towards non-atomic termination of 
transactions, or semantically incorrect transaction outcomes.

\end{abstract}
\vspace{-0.1in}
\noindent
{\bf Keywords}: Distributed Transaction, Two Phase Commit, 
Fault Tolerance, Byzantine Agreement, Web Services

\Section{Introduction}
\vspace{-0.1in}
The two-phase commit (2PC) protocol~\cite{gray:2pc} is a standard
distributed commit protocol~\cite{xopen/xa} for distributed transactions. 
The 2PC protocol is designed with the assumptions that the coordinator and the 
participants are subject only to benign faults, and the coordinator can be 
recovered quickly if it fails. Consequently, the 2PC protocol does not work 
if the coordinator is subject to arbitrary faults (also known as Byzantine 
faults~\cite{lamport:byz}) because a faulty coordinator might send conflicting 
decisions to different participants. Unfortunately, with more and more 
distributed transactions running over the untrusted Internet, driven by the 
need for business integration and collaboration, and enabled by the latest 
Web-based technologies such as Web services, it is a realistic threat that 
cannot be ignored.

This problem is first addressed by Mohan et al. in~\cite{Mohan:podc83} 
by integrating Byzantine agreement and the 2PC protocol. The basic idea is 
to replace the second phase of the 2PC protocol with a Byzantine agreement 
among the coordinator, the participants, and some redundant nodes within 
the root cluster (where the root coordinator resides). This prevents the 
coordinator from disseminating conflicting transaction outcomes to different 
participants without being detected. However, 
this approach has a number of deficiencies. First, it requires
all members of the root cluster, including participants, to reach a 
Byzantine agreement for each transaction. This would incur very high 
overhead if the size of the cluster is large. Second, it does not offer
Byzantine fault tolerance protection for subordinate coordinators or 
participants outside the root cluster. Third, it requires the participants
in the root cluster to know all other participants in the same cluster,
which prevents dynamic propagation of transactions. In general, only
the coordinator should have the knowledge of the participants set for each 
transaction. These problems prevent this approach from being used in 
practical systems.

Rothermel {\em et al.}~\cite{Rothermel:acmtds93} 
addressed the challenges of ensuring atomic distributed commit in open 
systems where participants (which may also serve as subordinate coordinators) 
may be compromised. However, \cite{Rothermel:acmtds93} assumes that 
the root coordinator is trusted, which limits its usefulness. 
Garcia-Molina {\em et al.}~\cite{Garcia-Molina:sigmod84} discussed the 
circumstances when Byzantine agreement is needed for distributed 
transaction processing. Gray~\cite{gray:byz2pc} compared the problems 
of distributed commit and Byzantine agreement, and provided insight on the 
commonality and differences between the two paradigms.

In this paper, we carefully analyze the threats to atomic commitment
of distributed transactions and evaluate strategies to mitigate such 
threats. We choose to use a Byzantine agreement algorithm only among
the coordinator replicas, which avoids the problems in~\cite{Mohan:podc83}.
An obvious candidate for the Byzantine agreement algorithm is the 
Byzantine fault tolerance (BFT) algorithm described 
in~\cite{bft-acm} because of its efficiency.
However, the BFT algorithm is designed to ensure totally ordered atomic
multicast for requests to a replicated stateful server. We made a number of 
modifications to the algorithm so that it fits the problem of atomic 
distributed commit. The most crucial change is made to the first phase 
of the BFT algorithm, where the primary coordinator replica 
is required to use a decision certificate, which is a collection of 
the registration records and the votes it has collected
from the participants, to back its decision on a transaction's outcome.
The use of such a certificate is essential to enable a correct
backup coordinator replica to verify the primary's proposal. This also
limits the methods that a faulty replica can use to 
hinder atomic distributed commit of a transaction. 

We integrated our Byzantine fault tolerant distributed commit (BFTDC) protocol 
with Kandula, a well-known open source distributed commit framework for 
Web services~\cite{apache:kandula}. The framework is an implementation 
of the Web Services Atomic Transaction Specification (WS-AT)~\cite{ws-at}. 
The measurements show that our protocol incurs only moderate runtime
overhead during normal operations.

\vspace{-0.1in}
\Section{Background}
\label{sec2}
\vspace{-0.1in}

\SubSection{Distributed Transactions}
\vspace{-0.1in}
A distributed transaction is a transaction that spans across multiple sites 
over a computer network. It should maintain the same ACID 
properties~\cite{gray:2pc} as a local transaction does. One of the most 
interesting issues for distributed transactions is how to guarantee atomicity, 
\ie either all operations of the
transaction succeed in which case the transaction commits, or none of
the operations is carried out in which case the transaction aborts.

The middleware supporting distributed transactions is often called 
transaction processing monitors (or TP monitors in short). One of the main
services provided by a TP monitor is a distributed commit service, which
guarantees the atomic termination of distributed transactions.
In general, the distributed commit service is implemented by the 2PC protocol,
a standard distributed commit protocol \cite{xopen/xa}. 

According to the 2PC protocol, a distributed transaction is modelled to 
contain one coordinator and a number of participants. A distributed 
transaction is initiated by one of the participants, which is referred to as
the initiator. The coordinator is created when the transaction
is activated by the initiator. All participants are required to register with
the coordinator when they get involved with the transaction. As the name
suggests, the 2PC protocol commits a transaction in two phases. During
the first phase (also called prepare phase), a request is disseminated by 
the coordinator to all participants so that they can prepare to commit the 
transaction. If a participant is able to commit the transaction, it 
prepares the transaction for commitment and responds with a ``prepared'' vote. 
Otherwise, it votes ``aborted''. When a participant responded with 
a ``prepared'' vote, it enters a ``ready'' state. Such a participant must 
be prepared to either commit or abort the transaction. A participant
that has not sent a ``prepared'' vote can unilaterally abort the transaction.
When the coordinator has received votes from every
participant, or a pre-defined timeout has occurred, it starts the second
phase by notifying the outcome of the transaction. The coordinator decides
to commit a transaction only if it has received the ``prepared'' vote from 
every participant during the first phase. It aborts the transaction otherwise.

\vspace{-0.1in}
\SubSection{Byzantine Fault Tolerance}
\vspace{-0.1in}
Byzantine fault tolerance refers to the capability of a system to tolerate
Byzantine faults. It can be achieved by replicating the server and 
by ensuring all server replicas receive the same input in the same order. The
latter means that the server replicas must reach an agreement on the input
despite Byzantine faulty replicas and clients. Such an agreement is often
referred to as Byzantine agreement~\cite{lamport:byz}.

Byzantine agreement algorithms had been too expensive to be practical
until Castro and Liskov invented the BFT algorithm mentioned 
earlier~\cite{bft-acm}. The BFT algorithm is executed by a set of 
$3f+1$ replicas to tolerate $f$ Byzantine faulty replicas. One of the 
replicas is 
designated as the primary while the rest are backups. The normal operation of 
the BFT algorithm involves three phases. During the first phase (called 
pre-prepare phase), the primary multicasts a pre-prepare message containing 
the client's request, the current view and a sequence number assigned to 
the request to all backups. A backup verifies the request 
message and the ordering information. If the backup accepts the message, 
it multicasts to all other replicas a prepare message containing the ordering 
information and the digest of the request being ordered. 
This starts the second phase, \ie the prepare phase. A replica waits until
it has collected $2f$ matching prepare messages from different replicas before
it multicasts a commit message to other replicas, which starts the third 
phase (\ie commit phase). The commit phase ends when a replica has received 
$2f$ matching commit messages from other replicas. At this point, the request 
message has been totally ordered and it is ready to be delivered to the 
server application. 

If the primary or the client is faulty, a Byzantine agreement on the
order of a request might not be reached, in which 
case, a new view is initiated, triggered by a timeout on the current view. 
A different primary is 
designated in a round-robin fashion for each new view installed. 

\vspace{-0.1in}
\Section{BFT Distributed Commit}
\label{sec3}

\vspace{-0.1in}
\SubSection{System Models}
\vspace{-0.1in}
We consider transactional client/server applications supported by an 
object-based TP monitor such as the WS-AT conformant 
framework~\cite{apache:kandula} used
in our implementation. For simplicity, we assume a flat distributed 
transaction model. We assume that for
each transaction, a distinct coordinator is created. The lifespan
of the coordinator is the same as the transaction it coordinates.

All transactions are
started and terminated by the initiator. The initiator also propagates
the transaction to other participants. The distributed commit
protocol is started for a transaction when a commit/abort request is received 
from the initiator. The initiator is regarded as a special participant. 
In later discussions we do not distinguish the initiator and other 
participants unless it is necessary to do so. 

When considering the safety of our distributed commit protocol, we use an 
asynchronous distributed system model. However, to ensure liveness, 
certain synchrony must be assumed. 
Similar to \cite{bft-acm}, we assume that the message transmission 
and processing delay has an asymptotic upper bound. This bound is 
dynamically explored in the adapted Byzantine agreement algorithm in that 
each time a view change occurs, the timeout for the new view is doubled.
 
We assume that the transaction coordinator runs separately from the 
participants, and it is replicated. For
simplicity, we assume that the participants are not replicated.
We assume that $3f+1$ coordinator replicas are available,
among which at most $f$ can be faulty during a transaction. There is
no limit on the number of faulty participants. Similar 
to \cite{bft-acm},
each coordinator replica is assigned a unique id $i$, where $i$ 
varies from $0$ to $3f$. For view $v$, the replica whose id $i$ satisfies 
$i=v \: mod \: (3f+1)$ would serve as the primary. The view starts from 0. 
For each view change, the view number is increased by one and a new primary 
is selected.

In this paper, we call a coordinator replica correct if it does not fail
during the distributed commit for the transaction under consideration,
\ie it faithfully executes according to the protocol prescribed from the
start to the end. However, we call a participant correct if it is not
Byzantine faulty, \ie it may be subject to typical non-malicious faults
such as crash faults or performance faults.

The coordinator replicas are subject to Byzantine faults, \ie a Byzantine
faulty replica can fail arbitrarily.
For participants, however, only a subset of faulty behaviors are tolerated,
such as a faulty participant sending conflicting votes to different 
coordinator replicas. Some forms of participant Byzantine behaviors
cannot be addressed by the distributed commit protocol.\footnote{
For example, a Byzantine faulty participant can vote to commit a 
transaction while actually aborting it, and vice versa.}

For the initiator, we further limits its Byzantine faulty behaviors.
In particular, it does not exclude any correct participant from the scope 
of the transaction, or include any participant that has not registered
properly with the coordinator replicas, as discussed below.

To ensure atomic termination of a distributed transaction, it is essential
that all correct coordinator replicas agree on the set of participants 
involved in the transaction. In this work,
we defer the Byzantine agreement on the participants set until the
distributed commit stage and combine it with that for the transaction
outcome. To facilitate this optimization, we need to make the following 
additional assumptions. 

We assume that there is proper authentication mechanism
in place to prevent a Byzantine faulty process from illegally
registering itself as a participant at correct coordinator 
replicas. Furthermore, we assume that a correct
participant registers with $f+1$ or more correct 
coordinator replicas before it sends a reply to the initiator when the
transaction is propagated to this participant with a request coming from
the initiator. If a correct participant crashes before the transaction 
is propagated to itself, or before it finishes registering with the
coordinator replicas, either no reply is sent back to the initiator,
or an exception is thrown back to the initiator. As a result, the initiator
should decide to abort the transaction. The interaction pattern among
the initiator, participants and the coordinator is identical to that
described in the WS-AT specification~\cite{ws-at}, except that the coordinator
is replicated in this work. 

All messages between the coordinator and the participants are 
digitally signed. We assume that the 
coordinator replicas and the participants each has a public/secret key pair. 
The public keys of the participants are known to all coordinator replicas,
and vice versa, while the private key is kept secret to its owner. 
We assume that the adversaries have limited computing 
power so that they cannot break the encryption and digital signatures
of correct coordinator replicas. 

\begin{figure}[t]
\centering
\includegraphics[width=2.85in]{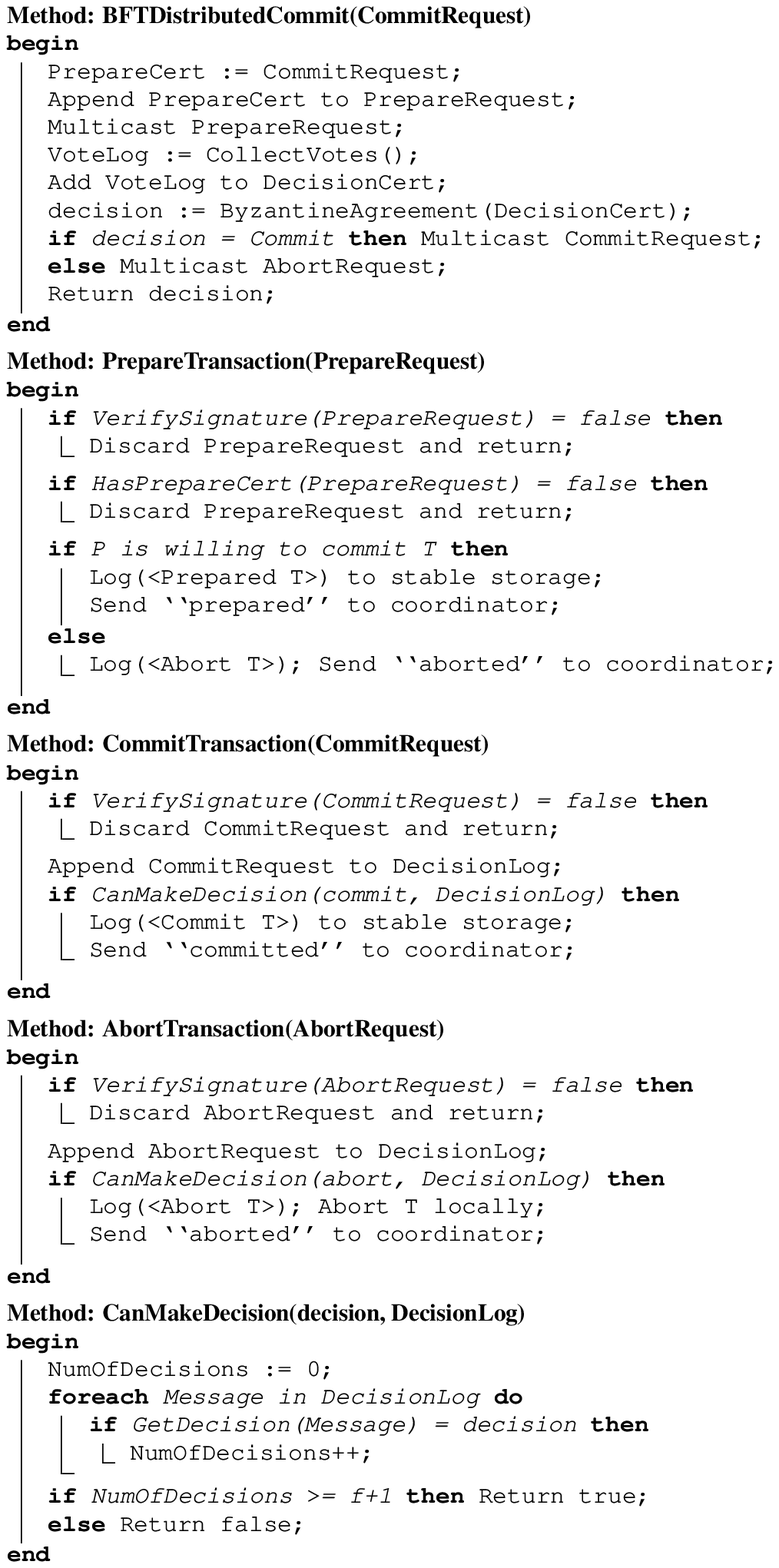}
\caption[]{Pseudo-code for our Byzantine fault tolerant distributed commit
protocol.}
\label{bftdcfig}
\vspace{-0.25in}
\end{figure}

\vspace{-0.1in}
\SubSection{BFTDC Protocol}
\vspace{-0.1in}
Figure~\ref{bftdcfig} shows the pseudo-code of the our Byzantine 
fault tolerant distributed commit protocol. Comparing with the 2PC
protocol, there are two main differences:
\begin{itemize}
\item[--] At the coordinator side, an additional phase of Byzantine agreement 
is needed for the coordinator replicas to reach a consensus
on the outcome of the transaction, before they notify the participants.
\vspace{-0.1in}
\item[--] At the participant side, a decision (commit or abort request)
from a coordinator replica is queued until at least f+1 identical
decision messages have been received, unless the participant unilaterally
aborts the transaction. This is to make sure that at least
one of the decision messages come from a correct coordinator replica.
\end{itemize}

The distributed commit for a transaction starts when a coordinator replica 
receives a commit request from the initiator. If the coordinator replica
receives an abort request from the initiator, it skips the first phase of the
distributed commit. In any case, a Byzantine agreement is conducted on the
decision regarding the transaction's outcome. 

The operations of each coordinator replica is defined in the 
BFTDistributedCommit() method in Fig.~\ref{bftdcfig}. During the prepare 
phase, a coordinator replica
sends a prepare request to every participant in the transaction. The
prepare request is piggybacked with a prepare certificate, which contains
the commit request sent (and signed) by the initiator.

When a participant receives a prepare request from a coordinator replica,
it verifies the correctness of the signature of the message and the
prepare certificate (if the participant does not know the initiator's
public key, this step is skipped). The prepare request is discarded
if any of the verification steps fails. Even though the check for a
prepare certificate is not essential to the correctness of our distributed
commit protocol, it nevertheless can prevent a faulty coordinator replica
from instructing some participants to prepare a transaction, even after
the initiator has requested to abort the transaction. 

At the end of the prepare phase, all correct coordinator
replicas engage in an additional round for them to reach a Byzantine 
agreement on the outcome of the transaction. The Byzantine agreement 
algorithm used in this phase is elaborated in Section~\ref{basec}.

When a participant receives a commit request from a coordinator replica,
it commits the transaction only if it has received the same decision
from $f$ other replicas so that at least one of them comes from 
a correct replica. The handling of an abort request is similar.

\vspace{-0.1in}
\SubSection{Byzantine Agreement Algorithm}
\label{basec}
\vspace{-0.1in}
The Byzantine agreement algorithm used in the BFTDC protocol is adapted from 
the BFT algorithm by Castro and Liskov~\cite{bft-acm}. To avoid 
possible confusion with the terms used to refer to the distributed commit 
protocol, the three phases during normal operations are referred to as 
ba-pre-prepare, ba-prepare, and ba-commit. Our algorithm differs from the 
BFT algorithm in a number of places due to different objectives. The BFT 
algorithm is used for server replicas to agree on the total ordering of the 
requests received, while our algorithm is used for the coordinator replicas 
to agree on the outcome (and participants set) of each transaction. 
In our algorithm,
the ba-pre-prepare message is used to bind a decision (to commit or abort) 
with the transaction under concern (represented by a unique transaction id).
In~\cite{bft-acm}, the ba-pre-prepare message is used to bind a 
request with an execution order (represented by a unique sequence number).
Furthermore, for distributed commit, an instance of our algorithm is
created and executed for each transaction. When there are multiple concurrent 
transactions, multiple instances of our algorithm are running concurrently and
independently from each other (the relative ordering of the distributed
commit for different transactions is not important). 
In~\cite{bft-acm}, however, a single instance of the BFT algorithm 
is used for {\em all} requests to be ordered.

When a replica completes the prepare phase of the distributed 
commit for a transaction, an instance of our Byzantine agreement algorithm 
is created. The algorithm starts with 
the ba-pre-prepare phase. During this phase, the primary $p$ sends a
ba-pre-prepare message including its decision certificate to all other 
replicas. The ba-pre-prepare message has the form 
$<${\sc ba-pre-prepare}$, v, t, o, C$$>_{\sigma_p}$,
where $v$ is the current view number, $t$ is the transaction id, 
$o$ is the proposed transaction outcome (\ie commit or abort),
$C$ is the decision certificate, and $\sigma_p$ is the signature of the 
message signed by the primary. The decision certificate contains a collection 
of records, one for each participant. The record for a participant $j$ 
contains a signed registration $R_j=(t, j)_{\sigma_j}$ and a signed vote
$V_j=(t, vote)_{\sigma_j}$ for the transaction $t$, if a vote from $j$ has been
received by the primary. The transaction id is included in each 
registration and vote record so that a faulty primary cannot reuse an 
obsolete registration or vote record to force a transaction outcome against 
the will of some correct participants 
(which may lead to non-atomic transaction commit). 

A backup accepts a ba-pre-prepare message provided:
\begin{itemize}
\item[--] The message is signed properly by the primary.
The replica is in view $v$, and it is handling transaction $t$.
\vspace{-0.06in}
\item[--] It has not accepted a ba-pre-prepared message for transaction $t$
in view $v$.
\vspace{-0.06in}
\item[--] The registration records in $C$ are identical to, or form a
superset of, the local registration records. 
\vspace{-0.06in}
\item[--] Every vote record in $C$ is properly signed by its sending 
participant and the transaction identifier in the record matches that of the 
current transaction, and the proposed decision $o$ is consistent with the 
registration and vote records.
\end{itemize}
Note that a backup does not insist on receiving a decision
certificate identical to its local copy. This is because a correct
primary might have received a registration from
a participant which the backup has not, or the primary and
backups might have received different votes from some Byzantine faulty
participants, or the primary might have received a vote that 
a backup has not received if the sending participant 
crashed right after it has sent its vote to the primary.

If the registration records in $C$ form a superset of the local registration
records, the backup updates its registration records and asks the 
primary replica for the endpoint reference\footnote{The term endpoint 
reference refers to the physical contact information such as host and
port of a process. In Web services, an endpoint reference typically contains a
URL to a service and an identifier used by the service to
locate the specific handler object~\cite{ws-addr}.} of each missing participant
(so that it can send its notification to the participant). 

A backup suspects the primary and initiates a view change
immediately if the ba-pre-prepare message fails the verification.
Otherwise, the backup accepts the ba-pre-prepare message.
At this point, we say the replica has {\em ba-pre-prepared} for 
transaction $t$. It then logs the accepted ba-pre-prepare message
and multicasts a ba-prepare message with the
same decision $o$ as that in the ba-pre-prepare message (this starts the 
ba-prepare phase). The ba-prepare message takes the form 
$<${\sc ba-prepare}$, v, t, d, o, i$$>_{\sigma_i}$,
where $d$ is the digest of the decision certificate $C$.

A coordinator replica $j$ accepts a ba-prepare message provided:
\begin{itemize} 
\item[--] The message is correctly signed by replica $i$, and
replica $j$ is in view $v$ and the current transaction is $t$;
\vspace{-0.06in}
\item[--] The decision $o$ matches that in the ba-pre-prepare message;
\vspace{-0.06in}
\item[--] The digest $d$ matches the digest of the decision certificate 
in the accepted ba-pre-prepare message.
\end{itemize}
If a replica has collected $2f$ matching ba-prepare messages from different 
replicas (including the replica's own ba-prepare message if it is a backup), 
the replica is said to have {\em ba-prepared} to make a decision on 
transaction $t$. This is the end of the ba-prepare phase.

A ba-prepared replica enters the ba-commit phase by multicasting a
ba-commit message to all other replicas. The ba-commit message has the
form $<${\sc ba-commit}$, v, t, d, o, i$$>_{\sigma_i}$. The replica $i$ 
is said to have {\em ba-committed}, if it has obtained
$2f+1$ matching ba-commit messages from different replicas (including
the message it has sent). 
When a replica is ba-committed for transaction $t$, it sends the decision 
$o$ to all participants of transaction $t$.

If a replica $i$ could not advance to the {\em ba-committed} state
until a timeout, it initiates a view change by sending
a view change message to all other replicas. The view change message
has the form $<${\sc view-change}$, v+1, t, P,i$$>_{\sigma_i}$, 
where $P$ contains information regarding its current state. If the replica 
has ba-pre-prepared $t$ in view $v$, it includes a tuple 
$<$$v, t, o, C$$>$. If it has ba-prepared $t$ in view $v$, it includes 
both the tuple $<$$v, t, o, C$$>$ and $2f$ matching ba-prepared messages from 
different replicas for $t$ obtained in view $v$. If the replica has 
not ba-pre-prepared $t$, it includes its own decision certificate $C$.

A correct replica that has not timed out the current view multicasts a view 
change message only if it is in view $v$ and it has received valid view 
change messages for view $v+1$ from $f+1$ different 
replicas. This is to prevent a faulty replica from inducing unnecessary 
view changes. A view change message is regarded as valid if it is for 
view $v+1$ and the ba-pre-prepare and ba-prepare information included in $P$, 
if any, is for transaction $t$ in a view up to $v$.

When the primary for view $v+1$ receives $2f+1$ valid view change 
messages for $v+1$ (including the one it has sent or would have sent), 
it installs the new view, and multicasts a new view message, in the
form $<${\sc new-view}$, v+1, V, t, o, C$$>$ for view $v+1$, where
$V$ contains $2f+1$ tuples for the view change messages received for 
view $v+1$. Each tuple has the form $<$$i, d$$>$, where $i$ is
the sending replica, and $d$ is the digest
of the view change message. The proposed decision $o$ for $t$ and the decision
certification $C$ are determined according to the following rules:
\begin{enumerate}
\item If the new primary has received a view change message containing a 
valid ba-prepare record for $t$, and there is no conflicting ba-prepare
record, it uses that decision.
\vspace{-0.03in}
\item Else, the new primary rebuilds a set of registration records from the
received view change messages. This new set may be identical to, or
a superset of, the registration set known to the new primary prior to
the view change. The new primary then rebuilds a set of vote records in
a similar manner. It is possible that conflicting vote records are found
from the same participant (\ie, a participant sent a ``prepared'' vote
to one coordinator replica, while sending an ``aborted'' vote to some
other replicas), in which case, a decision has to be made
on the direction of the transaction $t$. In this work, we choose to
take the ``prepared'' vote to maximize the commit rate. A new decision 
certificate will be constructed and a decision for $t$'s outcome is
proposed accordingly. They will be included in the new view message for view
$v+1$.
\end{enumerate}

When a backup receives the new view message, it verifies the
message basically by following the same steps used by the primary. If the
replica accepts the new view message, it may need to retrieve the
endpoint references for some participants that it did not receive
from other correct replicas. When a backup replica
has accepted the new view message and obtained all missing information, 
it sends a ba-prepare message to all other replicas. 
The algorithm then proceeds according to its normal operations.

\vspace{-0.1in}
\SubSection{Informal Proof of Correctness}
\vspace{-0.1in}
We now provide an informal proof of the safety of our Byzantine agreement
algorithm and the distributed commit protocol. Due to space limitation, the
proof for liveness is omitted.

{\em Claim 1: If a correct coordinator replica ba-commits a transaction $t$
with a commit decision, the registration records of all correct 
participants must have been included in the decision certificate, and
all such participants must have voted to commit the transaction.}

We prove by contradiction. Assume that there exists a correct participant 
$p$ whose
registration is left out of the decision certificate. Since a correct 
coordinator replica has ba-committed $t$ with a commit decision, it must have 
accepted a ba-pre-prepare message and 
$2f$ matching ba-prepare message from different replicas. 
This means that a set $R_1$ of $2f+1$ replicas have all accepted 
the same decision certificate without the participant $p$, the initiator
has requested the coordinator replicas to commit $t$, and every 
participant in the registration set has voted to commit the transaction. 
This further implies that the initiator
has received normal replies from all participants, including $p$, to which 
it has propagated the current transaction. Because the participant $p$ is
correct and responded to the initiator's request properly, it must have
registered with at least $2f+1$ coordinator replicas prior to sending
its reply to the initiator. Among the $2f+1$ coordinator replicas, at 
least a set $R_2$ of $f+1$ replicas are correct, \ie all replicas in
$R_2$ are correct and have the registration record for $p$
prior to the start of the distributed commit for $t$. Because the total
number of replicas is $3f+1$, the two sets $R_1$ and $R_2$ must intersect
in at least one correct replica. The correct replica in
the intersection either did not receive the registration from $p$, or
it has accepted a decision certificate without the registration record
for $p$ despite the fact that it has received the registration from $p$,
which is impossible. 
Therefore, all correct participants must have been included in the decision
certificate if any correct replica ba-committed a transaction with a commit
decision.

We next prove that if any correct replica ba-committed a transaction with 
a commit decision, all correct participants must have voted to commit the 
transaction.
Again, we prove by contradiction. Assume that the above statement is not
true, and a correct participant $q$ has voted to abort the transaction $t$. 
Since we have proved above that $q$'s registration record must have been 
included in the decision certificate, its vote 
cannot be ignored. Furthermore, since a correct replica ba-committed $t$ 
with a commit decision, the set $R_1$ of $2f+1$ replicas have
all accepted the commit decision. Again, since $R_1$ and $R_2$ must 
intersect by at least one correct replica, that replica both accepted
the commit decision and has received the ``aborted'' vote from $q$. This is
possible only if the ba-pre-prepare message that the replica has accepted 
contains a ``prepared'' vote from $q$. This contradict to the fact that
$q$ is a correct participant. A correct participant never sends conflicting
votes to different coordinator replicas. This concludes our proof for claim 1.

{\em Claim 2: Our Byzantine agreement algorithm ensures that all correct
coordinator replicas agree on the same decision regarding the outcome
of a transaction.}

We prove by contradiction. Assume that two correct replicas
$i$ and $j$ reach different decisions for $t$,
without loss of generality, assume $i$ decides to abort $t$ in a view
$v$ and $j$ decides to commit $t$ in a view $u$.

First, we consider the case when $v=u$. According to our algorithm, 
$i$ must have accepted a ba-pre-prepare message with an abort decision 
supported by a decision certificate, and $2f$ matching ba-prepare messages 
from different replicas, all in view $v$, this means a set $R_3$ of at least 
$2f+1$ replicas have ba-prepared $t$ with an abort decision 
in view $v$. Similarly, replica $j$ must have accepted a ba-pre-prepare 
message with a commit decision supported by a decision certificate, and 
$2f$ matching ba-prepare messages from different replicas for transaction 
$t$ in the same view $v$, which means a set $R_4$ of at least $2f+1$ replicas 
have ba-prepared $t$ with a commit decision in view $v$. Since there are 
only $3f+1$ replicas, the two sets $R_3$ and $R_4$ must intersect in at 
least $f+1$ replicas, among which, at least one is a correct replica. 
It means that this replica must have accepted two conflicting ba-pre-prepare
messages (one to commit and the other to abort) in the same view. 
This contradicts the fact that it is a correct replica.

Next, we consider the case when view $u>v$.
Since replica $i$ ba-committed with an abort decision for $t$ in view
$v$, it must have received $2f+1$ matching ba-commit messages from
different replicas (including the one sent by itself). This means that
a set $R_5$ of $2f+1$ replicas have ba-prepared $t$ in view $v$,
all with the same decision to abort $t$. To install a new view, the primary of
the new view must have received view change messages (including
the one it has sent or would have sent) from a set $R_6$ of $2f+1$ replicas.
Similar to the previous argument, the two sets $R_5$ and $R_6$ intersect
in at least $f+1$ replicas, among which, at least one must be a correct 
replica. This replica would have included the decision and the decision
certificate backed by the ba-pre-prepare message and the $2f$ matching 
ba-prepare messages it has received from other replicas, in its view change 
message. The primary in the new view, if it is correct, 
must have used the decision and decision certificate from this replica. 
This should have led all correct replicas to ba-commit transaction $t$ 
with an abort decision, which contradicts to the assumption that a correct 
replica committed $t$. If the primary is faulty and did not obey the new 
view construction rule, we argue that 
no correct replica could have accepted the new view message, let alone
to have ba-committed $t$ with a commit decision. Recall that a correct
replica should verify the new view message by following the new view
construction rules, just as a correct primary would do. We have proved above
that the $2f+1$ view change messages must contain one sent by a correct
replica with ba-prepare information for an abort decision. A correct replica
cannot possibly have accepted the new view message sent by the faulty primary,
which contains a conflicting decision. This contradicts to the initial
assumption that a correct replica $j$ committed transaction $t$ in view $u$.
The proof for the case when $v>u$ is similar.
Therefore, claim 2 is correct.

{\em Claim 3: The BFTDC protocol guarantees atomic termination of
transactions at all correct participants.}

We prove by contradiction. Assume that a transaction $t$ commits at
a participant $p$ but aborts at another participant $q$. According to
the criteria indicated in the CommitTransaction() method shown in
Fig.~\ref{bftdcfig}, $p$ commits the transaction $t$ only if it has
received the commit request from at least $f+1$ different coordinator 
replicas. Since at most $f$ replicas are faulty, at least one request 
comes from a correct replica.
Due to claim 1, if any correct replica ba-committed a transaction with a
commit decision, then the registration records of all correct 
participants must have been included in the decision certificate, and
all correct participants must have voted to commit the transaction. 

On the other hand, since $q$ aborted $t$, one of the following two 
scenarios must be true: (1) $q$ unilaterally aborted $t$, in which case, 
it must not have sent a ``prepared'' vote to any coordinator replica.
(2) $q$ received a prepare request, prepared $t$, sent a
``prepared'' vote to one or more coordinator replicas. But it received
an abort request from at least $f+1$ different coordinator replicas.

If the first scenario is true, $q$ might or might not have finished
its registration process. If it did not, the initiator would have been
notified by an exception, or would have timed out $q$. In any case, the
initiator should have decided to abort $t$. This conflicts
with the fact that $p$ has committed $t$ because it implies that the 
initiator has asked the coordinator replicas to commit $t$. If $q$ 
completed the registration process, its registration record should have
been aware by a set $R_7$ of at least $f+1$ correct replicas. 
Since $p$ has committed $t$, at least one correct replica has 
ba-committed $t$ with a commit decision, which in turn implies 
that a set $R_8$ of at least $2f+1$ coordinator replicas have 
accepted a ba-pre-prepare message with a decision certificate either has 
no $q$ in its registration records, or without $q$'s ``prepared'' vote. 
Since there are $3f+1$ replicas, $R_7$ and $R_8$ must intersect in at least one
replica. This correct replica could not possibly have accepted a ba-pre-prepare
message with a decision certificate described above.

\begin{figure*}[t]
\leavevmode
\hbox to \textwidth{
\vbox{
\hbox to 3.4in{\epsfxsize=3.4in \epsfbox{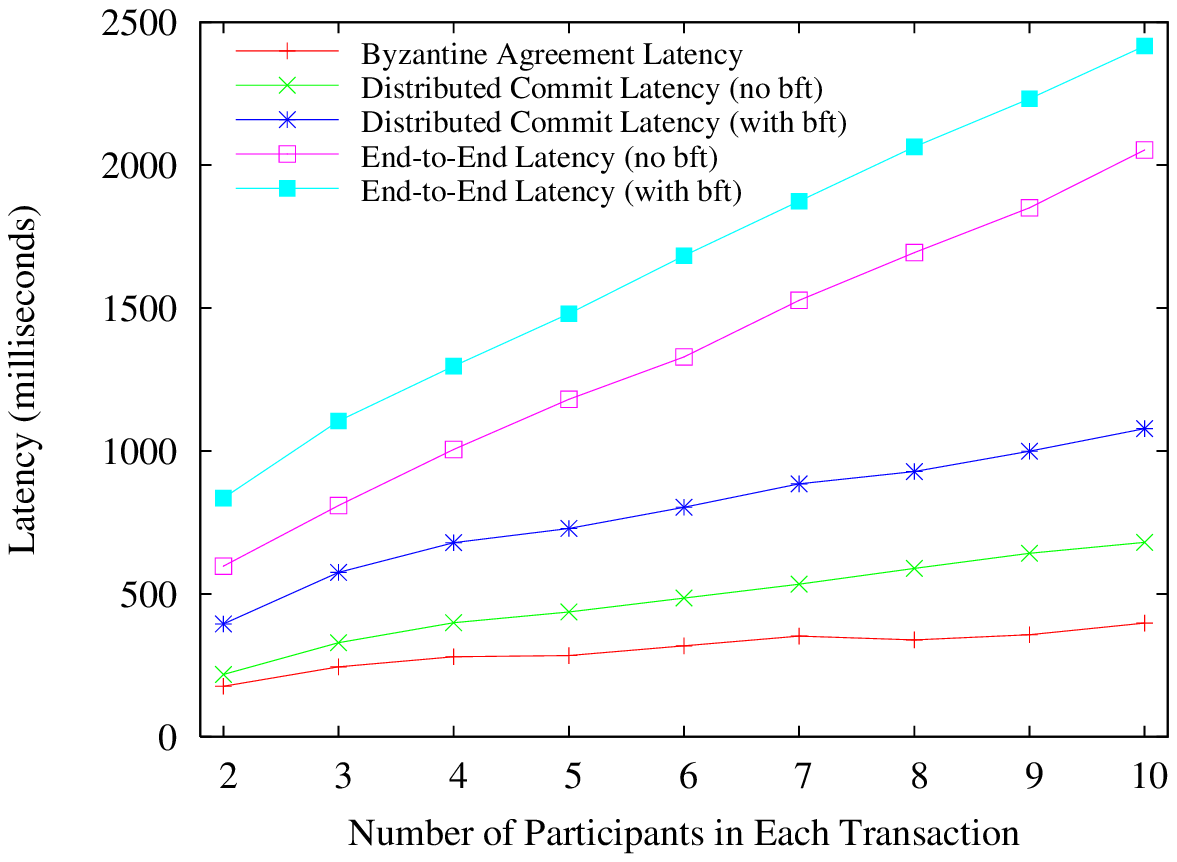}}
\hbox{\hspace{1.75in}(a)}
}
\hfil
\vbox{
\hbox to 3.4in{\epsfxsize=3.4in \epsfbox{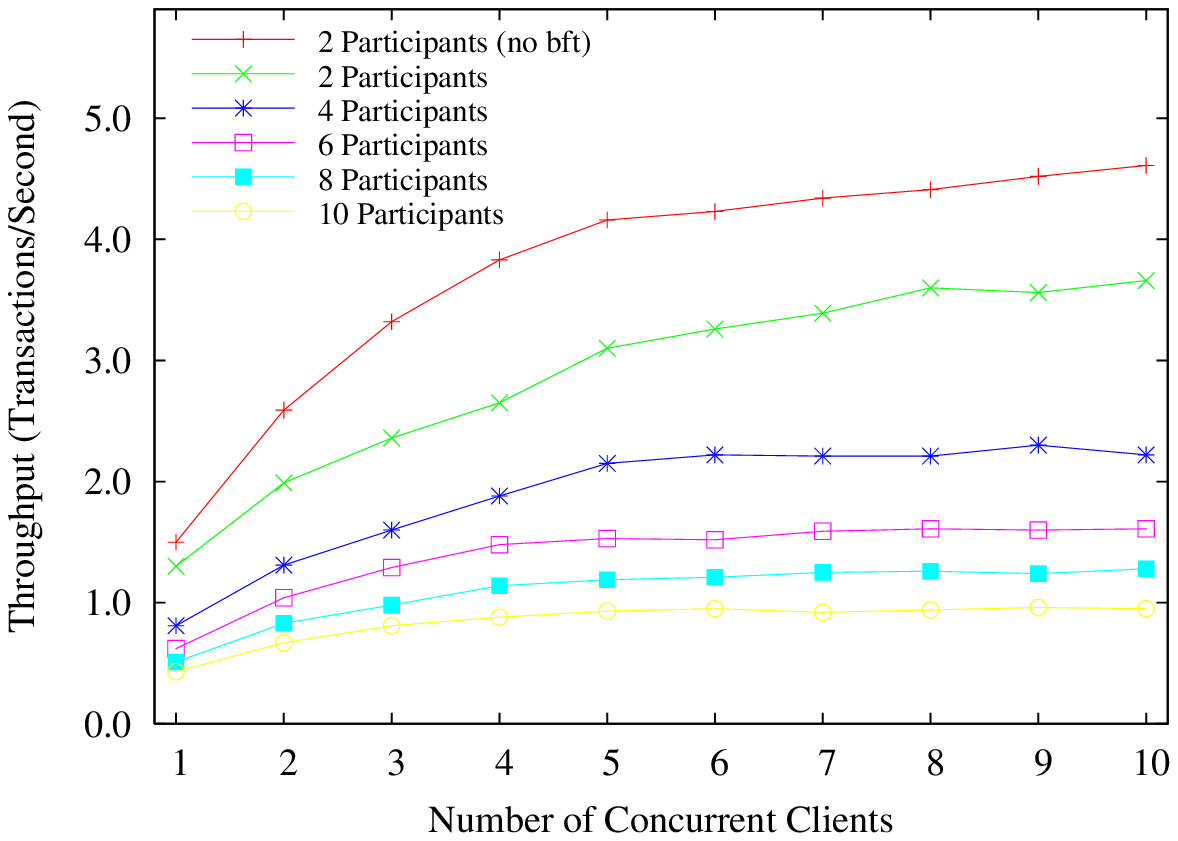}}
\hbox{\hspace{1.75in}(b)}
}}
\caption[]{(a) Various latency measurements for transactions with different
number of participants under normal operations (with a single client). 
(b) Throughput of the distributed commit service
in terms of transactions per second for transactions with different number
of participants under different load.}
\label{perffig}
\vspace{-0.25in}
\end{figure*}

For the second scenario, at least one correct replica has decided to abort 
$t$. Since another participant $p$ committed $t$, at least
one correct replica has decided to commit $t$. This contradicts to claim 2,
which we have proved to be true. Therefore, claim 3 is correct.

\vspace{-0.1in}
\Section{Implementation and Performance}
\label{sec4}
\vspace{-0.1in}
We have implemented the BFTDC protocol (with the exception of the view
change mechanisms) and integrated it into a distributed 
commit framework for Web services in Java programming language. 
The extended framework is based on a number of Apache Web services projects, 
including Kandula (an implementation of WS-AT) \cite{apache:kandula}, 
WSS4J (an implementation of the Web Services Security 
Specification) \cite{apache:wss4j}, and 
Apache Axis (SOAP Engine) \cite{axis}.
Most of the mechanisms are implemented in terms of 
Axis handlers that can be plugged into the framework without
affecting other components. Some of the Kandula code is modified 
to enable the control of its internal state, to enable a Byzantine agreement
on the transaction outcome, and to enable voting. Due to space constraint,
the implementation details are omitted.

For performance evaluation, we focus on assessing the runtime overhead of our 
BFTDC protocol during normal operations. Our experiment is carried
out on a testbed consisting of 20 Dell SC1420 servers connected
by a 100Mbps Ethernet. Each server is equipped with two 
Intel Xeon 2.8GHz processors and 1GB memory running SuSE 10.2 Linux. 

The test application is a simple banking Web services application where
a bank manager (\ie initiator) transfers funds among the participants within
the scope of a distributed transaction for each request received from a client.
The coordinator-side services are replicated on 4 nodes to tolerate a 
single Byzantine faulty replica. The initiator and other participants are not 
replicated, and run on distinct nodes. The clients are distributed
evenly (whenever possible) among the remaining nodes.
Each client invokes a fund transfer operation on the banking Web service 
within a loop without any ``think'' time between two consecutive calls.
In each run, 1000 samples are obtained. The end-to-end latency
for the fund transfer operation is measured at the client. 
The latency for the distributed commit and the Byzantine agreement is 
measured at the coordinator replicas.
Finally, the throughput of the distributed commit framework
is measured at the initiator for various number of participants and 
concurrent clients.

To evaluate the runtime overhead of our protocol, we compare the performance 
of our BFTDC protocol with the 2PC protocol as it is implemented in the 
WS-AT framework with the exception
that all messages exchanged over the network are digitally signed.

In Figure~\ref{perffig}(a), we included the distributed commit latency and 
the end-to-end latency for both our protocol (indicated by ``with bft'')
and the original 2PC protocol (indicated by ``no bft''). The Byzantine
agreement latency is also shown. Figure~\ref{perffig}(b) shows the
throughput measurement results for transactions using our protocol
with up to 10 concurrently running clients and 2-10 participants in each 
transaction. For comparison, the throughput for transactions using the
2PC protocol for 2 participants is also included.

As can be seen in Figure~\ref{perffig}(a), the latency for the distributed
commit and the end-to-end latency both are increased by about 200-400 ms 
when the number of participants varies from 2 to 10. This increase is
mostly attributed to the introduction of the Byzantine agreement phase in
our protocol. Percentage-wise, the end-to-end latency, as perceived by
an end user, is increased by only 20\% to 30\%, which is quite moderate.
We observe a similar range of throughput reductions for transactions using
our protocol, as shown in Figure~\ref{perffig}(b).

\vspace{-0.1in}
\Section{Conclusions}
\vspace{-0.1in}
\label{sec6}
In this paper, we presented a Byzantine fault tolerant distributed 
commit protocol. We carefully studied the types of Byzantine faults 
that might occur to a distributed transactional systems and identified
the subset of faults that a distributed commit protocol can handle.
We adapted Castro and Liskov's BFT algorithm to ensure Byzantine agreement
on the outcome of transactions. We also proved informally the correctness
of our BFTDC protocol. A working prototype of the protocol is built on top 
of an open source distributed commit framework for Web services. 
The measurement results of our protocol show only moderate runtime overhead. 
We are currently working on the implementation
of the view change mechanisms and exploring additional mechanisms to
protect a TP monitor against Byzantine faults, not only for 
distributed commit, but for activation, registration, and transaction
propagation as well.

\end{document}